\documentclass[aps,prl,twocolumn,byrevtex]{revtex4}
\usepackage{amsmath}

\usepackage{mathptmx}
\usepackage[scaled=.90]{helvet}
\usepackage{courier}

\newcommand\ie{i.e., } %
\newcommand\eg{e.g., } %
\newcommand\erf{\text{erf}} %
\newcommand\erfc{\text{erfc}} %
\newcommand\etal{\emph{et al.}} %
\newcommand\Lrt{\rho\sqrt{D t} } %
\newcommand\density{\overline{\rho}} %

\begin{document}

\title{Statistical properties of single-file diffusion front}

\author{Sanjib Sabhapandit} 

\affiliation{Laboratoire de Physique Th\'eorique et Mod\`eles Statistiques
  (UMR 8626 du CNRS), Universit\'e Paris-Sud, B\^atiment 100, 91405 Orsay
  Cedex, France}

\email{sanjib.sabhapandit@u-psud.fr}

\date{\today}

\begin{abstract} 
  Statistical properties of the front of a semi-infinite system of
  single-file diffusion (one dimensional system where particles cannot pass
  each other, but in-between collisions each one independently follow
  diffusive motion) are investigated. Exact as well as asymptotic results
  are provided for the probability density function of (a) the
  front-position, (b) the maximum of the front-positions, and (c) the
  first-passage time to a given position. The asymptotic laws for the
  front-position and the maximum front-position are found to be governed by
  the Fisher-Tippett-Gumbel extreme value statistics.  The asymptotic
  properties of the first-passage time is dominated by a
  stretched-exponential tail in the distribution.  The farness of the front
  with the rest of the system is investigated by considering (i) the gap
  from the front to the closest particle, and (ii) the density profile with
  respect to the front-position, and analytical results are provided for
  late time behaviors.
\end{abstract}


\maketitle

Diffusion of a set of labeled particles in the one-dimensional (1D) line
with the restriction that the particles cannot pass one another (so that
the order of particle labels remains same over time) is referred as
\emph{single-file diffusion} (SFD).  The notion of SFD was originally
introduced by Hodgkin and Keynes to describe the transport of ions through
narrow channels in biological membranes~\cite{Hodgkin}. Subsequently it has
been realized in a variety of systems including superionic or organic
conductors~\cite{Richards} and certain zeolites with a 1D pore
structure~\cite{zeolites}.  Interacting 1D particle systems, which include
the SFD, are of general interest~\cite{Spitzer, Liggett}, and have found
applications in diverse fields ranging from modeling collective dynamics of
molecular motors~\cite{Campas} to highway traffic flow~\cite{Chowdhury}.

The restriction on the motion of particles in SFD makes the movement of
individual particles (\emph{tagged-particles}) in the bulk anomalously
slow.  In an infinite system, the probability density function (PDF)
$p_0(x,t)$ of finding a particle at position $x$ at time $t$, initially
starting at the origin, approaches the Gaussian propagator at long time,
\begin{equation}
  p_0(x,t)\approx\frac{1}{\sqrt{2 \pi \langle x^2(t)\rangle}} 
  \exp\left(-\frac{x^2}{2 \langle x^2(t)\rangle}\right),
  \label{p0}
\end{equation}
where the mean-square displacement $\langle x^2(t)\rangle$ is proportional
to $\sqrt{t}$, rather than $t$ as found in the case of a simple diffusion
process.  This result was first obtained by Harris~\cite{Harris}, and
subsequently reestablished by various others using different
methods~\cite{Levitt, Fedders, Arratia, vanBeijeren, MajumdarBarma, Hahn,
  Rodenbeck, Kollmann}.

The anomalous mean-square displacement in SFD systems has been observed
experimentally in diffusion measurements made with pulse field gradient NMR
on methane or CF$_4$ confined in uni-dimensional channels in AlPO$_4$-5
zeolite~\cite{Gupta, Hahn2, Kukla}. Recent advances in experimental
techniques have made it possible to create well-defined SFD model-system by
confining paramagnetic colloidal spheres in trenches fabricated by
photolithography, where the particle-particle interaction can be precisely
adjusted by an external field~\cite{Wei}, or by confining charged colloidal
particles in 1D channels generated by scanning optical
tweezers~\cite{Lutz}.  These systems allow the trajectories of individual
particles to be followed over long period of time with video microscopy,
and hence it has been possible to experimentally verify the theoretical
result given by Eq.~\eqref{p0}, together with the anomalous mean-square
displacement~\cite{Wei}.

The theoretical and experimental studies in SFD systems, thus far, have
concerned with the properties of tagged-particles in the bulk. However, a
particle at the boundary should obey quite different laws from those of a
tagged-particle in the bulk.  The theoretical effort to understand those
laws has been missing thus far, which may be associated with the lack of an
corresponding ideal experimentally accessible model-system until recently.
However, with the present level of experimental sophistication~\cite{Wei,
  Lutz}, it would be feasible to have accurate measurements on a particle
at the boundary of a SFD model-system.  This motivates the present Letter
to initiate the needed theoretical study on the properties of the
\emph{front} of a semi-infinite SFD system.  One indeed finds that the
SFD-front obeys rather different and interesting laws, which would
encourage experimental measurements of new kind.  In a classical
single-file system, namely, the \emph{Jepsen gas}~\cite{Jepsen} --- the
particles undergo binary elastic collisions, but move ballistically
in-between collisions --- the front-velocity has been studied
recently~\cite{Bena}, which is also related to the \emph{fitness} in a
simple toy model of biological evolution of quasispecies~\cite{Krug, Jain}.
Therefore, the theoretical results of this Letter on the SFD-front would
also be of interest to this discipline.

Consider a semi-infinite system consists of hard-point Brownian particles,
initially at $t=0$ distributing on the semi-infinite real line
$(-\infty,0]$ according to a Poisson process having density
(\emph{intensity}) $\rho$, with the ``zero''-th particle locating at the
origin, ---\ie the particle-positions are: $x_0(0)=0$ and $x_{i+1}(0)\le
x_{i}(0)$, $i=0,\dotsc,\infty$, where the inter-particle gaps,
$l_i=x_i(0)-x_{i+1}(0),\, i\ge 0$, are independent and identically
distributed (i.i.d.)  according to the PDF $\psi(l_i)= \rho \exp(-\rho
l_i)$.  As this system evolves in time, each particle diffuses
independently having the same diffusion constant $D$, except the hard-core
repulsion forbids one particle to pass another one when two particles meet,
\ie each particle acts as a hard wall for the other.  Thus, the order of
particle labels is preserved over time, $x_{i+1}(t) \le x_i(t), \forall\,
t\ge0, i\ge0$.  The aim of this Letter is to provide a comprehensive study
of the \emph{front}, described by its position $x_0(t)$, as the above SFD
system evolves in time.  It obtains exact as well as asymptotic analytical
results for the following experimentally measurable quantities associated
with the front: (A) the propagator $p(x,t)$, ---the PDF of finding the
front $x_0(t)$ at position $x$ at time $t$, (B) $p_{\max}(z,t)$ ---the PDF
of maximum front-position $z=\max\bigl(\{x_0(\tau)\}, 0\le \tau\le
t\bigr)$, (C) $p_{\text{F}}(\tau,z)$ ---the PDF of first-passage time
$\tau$ of the front to a given position $z$ ($z\ge 0$), (D)
$\psi_0(\epsilon,t)$ ---the PDF of the gap $\epsilon(t)=x_0(t)-x_1(t)$,
between the front and the next particle, and (E) the density profile
$\density(\ell,t)$ as a function of the distance $\ell$, $\ell\ge 0$,
measured from the position of the front.

We begin by explaining the two main steps involve in the calculations. The
first step relates the SFD problem to a non-interacting one where Brownian
trajectories $X_j(t)$, $j=0,\dotsc,\infty$, can pass each
other~\cite{Harris}. In the non-interacting problem, whenever two particles
pass each other their order in the 1D line is interchanged, and hence, the
order of the particle labels is not preserved over time. The ordered SFD
trajectories $\{x_i(t)\}$ can be constructed by tracing along the parts of
non-interacting trajectories $\{X_j(t)\}$, keeping the order in the 1D line
fixed, ---one single trajectory $x_i(t)$ in the SFD may consists parts of
several non-interacting trajectories $\{X_j(t)\}$. Therefore, under this
mapping, the position of the SFD front is given by the position of the
rightmost particle in the non-interacting system, \ie
\begin{equation}
  x_0(t)=\max\bigl(X_0(t),X_1(t),\dotsc,X_{\infty}(t)\bigr), \; \forall t.
\label{x0}
\end{equation}
In the second step, one considers time evolution of $N+1$ non-interacting
Brownian particles, starting from an initial configuration at $t=0$ where
one particle is located at the origin and $N$ particles are independently
and uniformly distributed in the interval $[-L,0]$ in the real line.  In
this non-interacting system, the required quantities are calculated for the
rightmost particle, --- which is the front $x_0(t)$ in the SFD system, ---
averaging over all possible initial positions of $N$ particles in the
interval $[-L,0]$.  One finally takes the limit $N\rightarrow\infty$,
$L\rightarrow\infty$, while holding the particle density $\rho=N/L$ fixed,
to obtain the required quantities for the original semi-infinite SFD
system.  Note that, the probability $\Psi_{N,L}(l)=(1-l/L)^N$, that an
inter-particle gap is larger than $l$, in the limit $N\rightarrow\infty$,
$L\rightarrow\infty$, keeping $\rho=N/L$ fixed, converges to
$\Psi(l)=\exp(-\rho l)$, which is signature of a Poisson process.  Having
described the main steps, we now proceed with the calculations.

{\bfseries Propagator}.---
Consider $P(x,t)=\int_{-\infty}^x p(x_0,t)\, dx_0$, the probability that
front-position $x_0(t)$ is less than $x$ at time $t$. In the
non-interacting particle picture, according to Eq.~\eqref{x0}, this is
exactly the probability that at $t$, all the particle-positions
$\{X_i(t)\}$ are less than $x$. For a single Brownian particle with the
initial position $X_i(0)=y_i$, the probability that at $t$ the position
$X_i(t)$ is less than $x$, is $P_1(x,t|y_i)=1-\int_x^{\infty} \phi(X_i-y_i,
t)\, dX_i$, where
\begin{equation}
  \phi(x,t)=\frac{1}{\sqrt{4\pi D t}} 
  \exp\left(-\frac{x^2}{4 D t}\right),
\label{phi}
\end{equation}
is the single-particle Gaussian propagator. Explicitly,
\begin{equation}
  P_1(x,t|y_i)=1-\frac{1}{2}
  \erfc\left(\frac{x-y_i}{\sqrt{4 D t}}\right).
\label{P_1}
\end{equation}
Now averaging over the initial position of $y_i$ in $[-L,0]$ gives
$P_{1,L}(x,t)=(1/L)\int_{-L}^0 P_1(x,t|y_i)\, dy_i$.  Therefore, the
probability of $N$ such non-interacting particles at $t$ having positions
less than $x$, after averaging over their initial positions in $[-L,0]$, is
$P_{N,L}(x,t)= \bigl[P_{1,L}(x,t)\bigr]^N$.  Finally the limit
$N\rightarrow\infty$, $L\rightarrow\infty$, holding the particle density
$\rho=N/L$ fixed, leads to
\begin{align} 
  P_{N,L}(x,t) &=\left[ 1-\frac{1}{2L}\int_{-L}^0
    \erfc\left(\frac{x-y}{\sqrt{4 D t}}\right)\,dy\right]^N
  \notag\\
  &\xrightarrow[N/L=\rho]{N,L\rightarrow\infty}
  \exp\Biggl(-\frac{\Lrt}{\sqrt{\pi}}\, S\biggl(\frac{x}{\sqrt{4 D
      t}}\biggr) \Biggr),
\label{limit P_{N,L}}
\end{align}
in which,
\begin{equation}
  S(u)=
  \exp(-u^2) -\sqrt{\pi}\,u\,\erfc(u).
  \label{S(u)}
\end{equation}
The desired probability $P(x,t)$ is, therefore, obtained by multiplying
Eq.~\eqref{limit P_{N,L}} by $P_1(x,t|0)$, for the particle that started
from the origin. Explicitly,
\begin{equation}
  P(x,t)=F\left(\frac{x}{\sqrt{4 D t}},t\right),
  \label{P(x,t)}
\end{equation}
where the scaling function is,
\begin{equation}
  F(u,t)=
  \left[1-\frac{\erfc(u)}{2}\right]
  \exp\left(-\frac{\Lrt}{\sqrt{\pi}}\, S(u) \right).
  \label{F}
\end{equation}
Equation~\eqref{P(x,t)}, together with Eqs.~\eqref{F} and \eqref{S(u)},
gives the exact properties of the SFD front for all $\Lrt$. The limit
$\rho=0$ recovers the result for a single diffusive particle.  The
propagator of the front is given by, $p(x,t)=\partial P(x,t)/\partial x$.
For $\Lrt \gg 1$, using the asymptotic form $S(u)\sim \exp(-u^2)/2 u^2$ in
Eq.~\eqref{F}, one finds the limiting form
\begin{equation}
  F(u,t) \approx
  G\left(\frac{u-a\bigl(\Lrt\big/2\sqrt{\pi}\bigr)} 
    {b\bigl(\Lrt\big/2\sqrt{\pi}\bigr)}\right),
\label{Limit F}
\end{equation}
in which the scaling parameters are,
\begin{subequations}
\begin{align}
  &a(c)= \sqrt{\ln c} -\ln \ln c/2\sqrt{\ln c}+\dotsb,\\
  &b(c)=1/2\sqrt{\ln c}+\dotsb,
\end{align}
\end{subequations}
and the scaling function is the Fisher-Tippett-Gumbel (FTG) distribution
function~\cite{EVS},
\begin{equation}
  G(x)=\exp\bigl(-\exp(-x)\bigr), 
  \quad -\infty\le x \le \infty.
\label{gumbel}
\end{equation}
Equivalent limiting result has been obtained by Arratia~\cite{Arratia}.
However, we have arrived at the result here more transparent way from the
exact result Eq.~\eqref{P(x,t)}.
The propagator of the front $p(x_0,t)$ for $\Lrt\gg1$, after the
appropriate scaling $x=\bigl[x_0/\sqrt{4 D
  t}-a\bigl(\Lrt\big/2\sqrt{\pi}\bigr)
\bigr]\big/b\bigl(\Lrt\big/2\sqrt{\pi}\bigr)$, is given by the FTG PDF
$g(x)=\exp\bigl(-x-\exp(-x)\bigr)$.

{\bfseries Maximum front-position}.---
Let $Q(z,t)$ be the probability that the front of the SFD system, with the
initial condition $x_0(0)=0$ and $-\infty< x_{i+1}(0) \le x_i(0) \le 0,
i=1,\dotsc,\infty$, does not cross the position $z$ ($z\ge 0$), up to time
$t$. Clearly, $Q(z,t)=\text{Prob}\bigl[\max\bigl(\{x_0(\tau)\}, 0\le
\tau\le t\bigr)<z\bigr]$, and the PDF of the maximum of the front-positions
in the time interval $[0,t]$ is given by $p_{\max}(z,t)=\partial
Q(z,t)/\partial z$. In the non-interacting picture, $Q(z,t)$ is exactly the
probability that none of the particle crosses $z$ up to $t$ [cf.
Eq.~\eqref{x0}]. Now, the probability that a Brownian particle starting
from $y$ ($y\le0$) at $t=0$ does not cross $z$ ($z\ge 0$) up to $t$
is~\cite{Majumdar, Redner},
\begin{math}
  Q_1(z,t|y)=\erf\bigl([z-y]\big/\sqrt{4 D t} \bigr),
\end{math}
known as the \emph{survival probability} in the literature.  Averaging over
the initial position in $[-L,0]$ gives $Q_1(z,t)=(1/L)\int_{-L}^0
Q_{1,L}(z,t|y)\,dy$. Then for $N$ non-interacting particles with initial
positions averaged in $[-L,0]$, the probability that none of them crosses
$z$ up to $t$ is simply $Q_{N,L}(z,t)=\bigl[Q_1(z,t)\bigr]^N$.  Finally the
limit $N\rightarrow\infty$, $L\rightarrow\infty$, holding $\rho=N/L$ fixed,
leads to [cf. Eq.~\eqref{limit P_{N,L}}]
\begin{equation}
  Q_{N,L}(z,t)
  \xrightarrow[N/L=\rho]{N,L\rightarrow\infty}
  \exp\Biggl(-\frac{2\Lrt}{\sqrt{\pi}}\, 
    S\biggl(\frac{z}{\sqrt{4 D t}}\biggr) \Biggr).
\end{equation}
Similar expression with $z=0$, \ie $S(0)=1$, has been encountered
elsewhere, \eg see Ref.~\cite{Bray}.  However, our interest lies in the
behavior of $Q(z,t)$ for $z\ge0$, which is exactly the product of the above
expression and the probability $Q_1(z,t|0)$, that the particle starting
from the origin does not cross $z$ in the interval $[0,t]$:
\begin{equation}
  Q(z,t)=\Omega\left(\frac{z}{\sqrt{4 D t}},t\right),
  \label{Q}
\end{equation}
where the exact scaling function is,
\begin{equation}
  \Omega(u,t)=
  \erf(u)\:
  \exp\left(-\frac{2\Lrt}{\sqrt{\pi}}\, S(u) \right).
\end{equation}
For $\Lrt\gg1$, $\Omega(u,t)$ has the same limiting FTG distribution as in
Eq.~\eqref{Limit F}, however, now at different values of the scaling
parameters, namely,
\begin{equation}
  \Omega(u,t) \approx
  G\left(\frac{u-a\bigl(\Lrt\big/\sqrt{\pi}\bigr)} 
    {b\bigl(\Lrt\big/\sqrt{\pi}\bigr)}\right).
  \label{Limit Omega}
\end{equation}
Therefore, for $\Lrt\gg 1$, the PDF of the maximum position
$p_{\max}(z,t)$, after the appropriate scaling, is given by the FTG PDF
$g(x)=\exp\bigl(-x-\exp(-x)\bigr)$.

{\bfseries First-passage time}.---
The time $\tau$ at which the SFD front crosses a specified position $z$ for
the first time, is the first-passage time (FPT) of the front to $z$.  The
FPT problems have wide implications and therefore, are of general
interest~\cite{Majumdar, Redner}. The probability $\int_t^\infty
p_{\text{F}}(\tau,z)\,d\tau$, that the FPT to $z$ ($z\ge 0$) is greater
than $t$, is exactly the probability that the SFD front has not cross $z$
in the interval $[0,t]$, which is given by $Q(z,t)$ in Eq.~\eqref{Q}.
Therefore, the PDF of the FPT to a given position $z$ is
$p_{\text{F}}(\tau,z)=-\partial Q(z,\tau)/\partial\tau$, with proper
normalization $\int_0^\infty p_{\text{F}}(\tau,z)\,d\tau=Q(z,0)=1$.  It is
convenient to express the PDF of the FPT in the form
\begin{equation}
  p_{\text{F}}(\tau,z) = A(\tau,z)\: \exp\bigl(-\Upsilon(\tau,z)\bigr),
\end{equation}
in which the two functions $A(\tau,z)$ and $\Upsilon(\tau,z)$ are,
\begin{align}
  &A(\tau,z)= \frac{z}{\sqrt{4 \pi D}\, \tau^{3/2}} +
  \frac{\rho\sqrt{D}}{\sqrt{\pi \tau}}
  \,\erf\left(\frac{z}{\sqrt{4 D \tau}}\right),\\
  &\Upsilon(\tau,z)=\frac{z^2}{4 D \tau} +\frac{2\rho\sqrt{
      D\tau}}{\sqrt{\pi}}\,S\left(\frac{z}{\sqrt{4 D \tau}}\right).
\end{align}
The limit $\rho=0$ reproduces the known result for the FPT of a single
Brownian particle, as expected.  For $\rho\not=0$, while $\tau\rightarrow
0$ limit is still governed by the essential singular behavior of the single
particle, at long time the asymptotic behavior of $p_{\text{F}}(\tau,z)$ is
dominated by the stretched-exponential tail
$\exp(-2\rho\sqrt{D\tau}/\sqrt{\pi})$, in contrast to the $\tau^{-3/2}$
power-law behavior of the single particle.

Finally, consider the question: \emph{How far is the front at a later time,
  away from the rest of the system?}  There is, of course, no unique way to
measure it quantitatively.  The remaining part of this Letter addresses the
question by considering two measurable quantities, namely, the gap between
the front and the next particle $\epsilon(t)=x_0(t)-x_1(t)$, and the
density profile with respect to the front.

{\bfseries Gap}.--- 
Consider $\Psi_0(\epsilon,t)=\int_\epsilon^\infty \psi_0(\epsilon',t)\,
d\epsilon'$, the probability that at time $t$ the gap is greater than
$\epsilon$.  Let $R(\epsilon,t|x)\, dx$ be the conditional probability that
the gap is greater than $\epsilon$, given that the front $x_0(t)$ lies
within $[x+\epsilon, x+\epsilon +dx]$.  Clearly,
$\Psi_0(\epsilon,t)=\int_{-\infty}^{\infty} R(\epsilon,t|x)\, dx$.
Let us use the notation $R_{N,L}(\epsilon,t|x)$ to refer $R(\epsilon,t|x)$,
in the finite non-interacting system, where initially one particle is
located at the origin, and $N$ others uniformly distributed in $[-L,0]$.
Then, in the non-interacting picture, $R_{N,L}(\epsilon,t|x)\, dx$ is given
by the probability that at $t$, one particle is located in $[x+\epsilon,
x+\epsilon +dx]$, and rest $N$ particles have positions less than $x$.  The
particle in $[x+\epsilon, x+\epsilon +dx]$ at $t$ is either the one
initially located at the origin or one out of $N$ particles initially
uniformly distributed in $[-L,0]$ ---the corresponding PDF is
$q_{N,L}(x+\epsilon,t)=(N/L)\int_{-L}^0 \phi(x+\epsilon-y_i,t)\,dy_i$.
Therefore,
\begin{math}
  R_{N,L}(\epsilon,t|x)= \phi(x+\epsilon,t) P_{N,L}(x,t) +
  q_{N,L}(x+\epsilon,t) P_1(x,t|0) P_{N-1,L}(x,t),
\end{math}
In the limit $N\rightarrow\infty$, $L\rightarrow\infty$, holding $\rho=N/L$
fixed, $q_{N,L}(x,t)\rightarrow (\rho/2)\, \erfc\bigl(x\big/\sqrt{4 D
  t}\bigr)$.  Thus, taking this thermodynamic limit in
$R_{N,L}(\epsilon,t|x)$ yields,
\begin{math}
  R(\epsilon,t|x) = W\bigl(\epsilon/\sqrt{4 D t},x/\sqrt{4 D t},
  t\bigr)/\sqrt{4 D t} ,
\end{math}
where [cf.  Eqs.~\eqref{phi}\nobreakdash--\eqref{limit P_{N,L}} and
\eqref{F}],
\begin{align}
  W(\delta, u, t)=& 
  \frac{1}{\sqrt{\pi}}\;  \exp\left(-(u+\delta)^2 
      -\frac{\Lrt}{\sqrt{\pi}}\, S(u) \right)\notag \\
&+ \Lrt\; \erfc(u+\delta)\;  F(u,t).
\label{W}
\end{align}
The desired probability for the gap can be obtained by integrating
$R(\epsilon,t|x)$ over $x$, ---\ie
\begin{math}
  \Psi_0(\epsilon,t)=\int_{-\infty}^{\infty} W\bigl(\epsilon\big/\sqrt{4 D
    t},u, t\bigr)\, du.
\end{math}
For $\Lrt\gg 1$, the main contribution comes from the second term of
Eq.~\eqref{W}. Using the limiting form of $F(u,t)$ given by Eq.~\eqref{F}
and the asymptotic form of $\Lrt\,\erfc(x)$, one finally obtains
$\Psi_0(\epsilon,t)\approx\exp\bigl(-\density_0(t)\, \epsilon\bigr)$, for
$\Lrt \gg 1$. Here $\density_0$ is the effective density near the front,
given by
\begin{equation}
  \density_0(t)
  =\left[\frac{\ln \bigl(\Lrt\big/2\sqrt{\pi}\bigr)}{D t} \right]^{1/2}.
  \label{rho_0}
\end{equation}

{\bfseries Density profile}.--- 
The density of particles at a position $x(t)$, is the PDF of finding a
particle at position $x$ at time $t$, is $(\rho/2)\,
\erfc\bigl(x\big/\sqrt{4 D t}\bigr) +\phi(x,t)$. However, the goal here is
to compute the density $\density(\ell,t)$ at a distance $\ell$, measured
from the front-position $x_0(t)$, ---\ie the PDF of finding a particle at a
distance $\ell$ away from the front-position.  Let $\density(\ell,t|x)$ be
the conditional PDF of finding a particle at position $x-\ell$, $\ell\ge
0$, given that the front is at position $x$.  Clearly, $\density(\ell,t)
=\int_{-\infty}^{\infty} \density(\ell,t|x)\, dx$.  In the non-interacting
picture, $\density(\ell,t|x)$ is the joint PDF of finding one particle at
position $x-\ell$, another one at $x$ and rest of them having positions
less than $x$. Taking the thermodynamic limit in the non-interacting
picture, one gets, $\density(\ell,t|x)=q_{\lim}(x-\ell,t)\, q_{\lim}(x,t)\,
P(x,t) + \dotsb$, where $q_{\lim}(x,t)=(\rho/2)\, \erfc\bigl(x\big/\sqrt{4
  D t}\bigr)$, is the PDF of finding a particle at position $x$ at time $t$
starting from the interval $(-\infty,0)$, and $P(x,t)$ is given by
Eq.~\eqref{P(x,t)}.  The remaining terms in $\density(\ell,t|x)$ are the
contributions of the events where the trajectories of the particle starting
from the origin reach $x$ and $x-\ell$ respectively, and can be neglected
for large $\Lrt$.  Therefore, integrating $\density(\ell,t|x)$ over $x$,
for $\Lrt\gg 1$,
\begin{equation}
  \density(\ell,t)\approx\frac{\rho}{2}\int_{-\infty}^{\infty} 
  \erfc\left(u-\frac{\ell}{\sqrt{4 D t}}\right) h(u,t)\;du,
\label{VR integral}
\end{equation}
where $h(u,t)=\Lrt\:\erfc(u)\, F(u,t)$. In the large $\Lrt$ limit,
$h(u,t)$, in terms of the scaled variable $x=\bigl[u-a(c)\bigr]\big/b(c)$
with $c=\Lrt\big/2\sqrt{\pi}$, converges to the FTG PDF,
$g(x)=\exp\bigl(-x-\exp(-x)\bigr)$.  Therefore, in the limit
$\Lrt\rightarrow\infty$, as $b\bigl(\Lrt\big/2\sqrt{\pi}\bigr)\rightarrow
0$, $h(u,t)$ becomes Dirac-delta function at
$u=a\bigl(\Lrt\big/2\sqrt{\pi}\bigr)$, and hence, Eq.~\eqref{VR integral}
yields
\begin{equation}
  \density(\ell,t)\approx \frac{\rho}{2}\;
  \erfc\left(a\biggl(\frac{\Lrt}{2\sqrt{\pi}}\biggr)
    -\frac{\ell}{\sqrt{4 D t}}\right).
  \label{density profile}
\end{equation}
Analogous limiting form also arises for the density of \emph{near-extreme}
events in i.i.d. random variables~\cite{Sabhapandit}.  In the
$\ell\rightarrow\infty$ limit, $\density(\infty,t)=\rho$, as expected.  For
$\ell=0$, using the asymptotic form $\erfc(a)\sim \exp(-a^2)/a\sqrt{\pi}$,
one recovers Eq.~\eqref{rho_0} with $\density(0,t)=\density_0(t)$.

In conclusion, this Letter has provided a comprehensive study of
statistical properties of the front of a semi-infinite single-file
diffusive system.  It would be interesting to investigate the effect of
bias, finite-ranged inter-particle interactions, and the disorder presence
in the system.  The ultimate interest, however, lies in realizing the
results in experimental systems.

\begin{acknowledgments}
  The author thanks S. N. Majumdar for useful comments and suggestions on
  the manuscript, and acknowledges the support of the Indo-French Centre
  for the Promotion of Advanced Research (IFCPAR/CEFIPRA) under Project
  3404-2.
\end{acknowledgments}

\end{document}